\documentclass[%
 aip,
 jmp,%
 amsmath,amssymb,
 reprint,%
]{revtex4-1}
\usepackage{graphicx}% Include figure files
\usepackage{dcolumn}% Align table columns on decimal point
\usepackage[version=3]{mhchem} % Formula subscripts using \ce{}
\usepackage{bm}% bold math

\usepackage[english]{babel}

\begin{document}

%\preprint{AIP/123-QED}

\title[Comparing the Accuracy of High-Dimensional Neural Network Potentials and the Fragmentation Method]{Comparing the Accuracy of High-Dimensional Neural Network Potentials and the Systematic Molecular Fragmentation Method: A Benchmark Study for all-trans Alkanes}

\author{M. Gastegger}
\affiliation{ 
Institute of Theoretical Chemistry, University of Vienna, W\"ahringer Stra{\ss}e 17, Vienna, Austria
}
\author{C. Kauffmann}
\affiliation{ 
Institute of Theoretical Chemistry, University of Vienna, W\"ahringer Stra{\ss}e 17, Vienna, Austria
}
\author{J. Behler}
\affiliation{ 
Lehrstuhl f\"ur Theoretische Chemie, Ruhr-Universit\"at Bochum, Universit\"atsstra{\ss}e 150, Bochum, Germany
}
\author{P. Marquetand}
\email{philipp.marquetand@univie.ac.at}
\affiliation{ 
Institute of Theoretical Chemistry, University of Vienna, W\"ahringer Stra{\ss}e 17, Vienna, Austria
}%

\date{\today}

\begin{abstract}

Many approaches, which have been developed to express the potential energy of large systems, exploit the locality of the atomic interactions. A prominent example are fragmentation methods, in which quantum chemical calculations are carried out for overlapping small fragments of a given molecule that are then combined in a second step to yield the system's total energy. Here we compare the accuracy of the systematic molecular fragmentation approach with the performance of high-dimensional neural network (HDNN) potentials introduced by Behler and Parrinello. HDNN potentials are similar in spirit to the fragmentation approach in that the total energy is constructed as a sum of environment-dependent atomic energies, which are derived indirectly from electronic structure calculations. As a benchmark set we use all-trans alkanes containing up to eleven carbon atoms at the coupled cluster level of theory. These molecules have been chosen because they allow to extrapolate reliable reference energies for very long chains, enabling an assessment of the energies obtained by both methods for alkanes including up to 10 000 carbon atoms.
We find that both methods predict high-quality energies with the HDNN potentials yielding smaller errors with respect to the coupled cluster reference. 

\end{abstract}

\maketitle

\section{Introduction}

Computer simulations of chemical processes rely on the potential energy surfaces (PESs) of the structures involved \cite{Levine2013}, and consequently the accuracy of these PESs defines the quality of the simulations. While highly accurate \textit{ab initio} calculations are at hand for moderately sized systems, larger systems can only be addressed by employing an increasing number of empirical approximations in order to keep the computational effort feasible, which necessarily results in a reduced accuracy of the obtained energies. Thus, maintaining accuracy while enabling a fast evaluation is one of the main goals when constructing PESs. Many different approaches have been developed in past decades, which have either been based on physical considerations or on purely mathematical principles. 

Within the latter subgroup, PESs derived from machine learning techniques~\cite{P4263}, and in particular employing neural networks (NNs)~\cite{blank_neural_1995,behler_generalized_2007,behler_representing_2014,manzhos_random-sampling_2006,manzhos_using_2006,pukrittayakamee_simultaneous_2009,Jiang_2013,nguyen_modified_2012,NIPS2012_0223,Zhang_2015,houlding_polarizable_2007}, have made a lot of progress. 
NNs are nonlinear models inspired by the central nervous system, which are especially adept at interpolating trends in existing data.
Their flexible and unbiased nature has lead to a variety of NN-based applications in many fields\cite{schmidhuber_2015} and makes them a useful tool for fitting PESs for different types of chemical systems~\cite{handley_potential_2010,behler_neural_2011}. 
However, early NN potentials usually required system-specific adoptions and were limited to small numbers of atoms, which has been finally resolved in the high-dimensional NN (HDNN) approach by Behler and Parrinello \cite{behler_generalized_2007}.

Still, the applicability of NN-based methods is limited by the need for large sets of \textit{ab initio} reference calculations in order to construct a valid and accurate potential.
Especially for large molecular systems -- such as proteins -- these reference calculations quickly become prohibitive, due to the scaling behavior of high-level \textit{ab initio} methods.
In the HDNN approach the need for reference calculations comprising the full systems of interest is circumvented by the exploitation of so-called\cite{He2014} chemical locality.
Consequently, it is possible to construct HDNNs based solely on fragments of the original molecular system, while the validity for the full system is retained.
Hence, one costly reference computation can be replaced by several significantly cheaper calculations on smaller subsystems.
This approach is well tested for solid state systems~\cite{behler_metadynamics_2008,P3114,artrith_neural_2013} as well as for molecular clusters~\cite{P3875} and liquid water~\cite{Morawietz-submitted}, and has been used in numerous applications~\cite{behler_representing_2014}.

Amongst the physically motivated approaches are fragmentation-based methods, where the original system is first divided into smaller independent subsystems. The properties of these fragments (e.g. energies) are then calculated with \emph{ab-initio} methods and recombined to obtain the composite properties of the whole molecular system. 
Several fragmentation schemes have been developed over the last 20 years, differing mainly in how the original system is divided into fragments and how the recombination step is carried out.\cite{Gordon2012,Collins_2015}
Here, we focus on the systematic molecular fragmentation approach (SMF) developed by Collins and coworkers.\cite{Collins2006,Netzloff2007,C2CP23832B}
The SMF approach generates overlapping fragments of a certain size by gathering bonded atoms into functional groups. The energy of the total system is calculated by summing the energies of these fragments and subtracting the energy contributions of the overlap regions.
SMF has been applied successfully to a wide range of molecular systems, including proteins\cite{C2CP23832B,Collins2014}, water clusters\cite{Collins2014}, \ce{SiO2} crystals\cite{Netzloff2007} and organic molecules\cite{Collins2009}.

The aim of the present study is to assess and compare the performance of the HDNN method and of the SMF approach in terms of the accuracy of the obtained potential energies.
For this purpose linear all-trans alkane chains of varying lengths containing up to 10~000 carbon atoms have been chosen as  model system. Reference computations for the shorter chains containing up to eleven carbon atoms have been carried out directly with the coupled cluster method including single, double and perturbative triple excitations (CCSD(T)), while the reference energies of longer chains have been extrapolated using corrected energies of the fragmentation approach.
Based on these reference calculations, we investigate how well HDNNs and the conventional fragmentation approach can predict the potential energies of large organic molecules if only the energies of the small fragments accessible by CCSD(T) are provided. The simplicity of our model system is motivated by the need for high-quality reference energies for very large molecules, which could not be obtained for more complex systems as detailed below, but our findings are general and not restricted to linear alkanes.

\section{Methods}

\subsection{High Dimensional Neural Network Potentials}

\begin{figure}[h]
  \centering
  \includegraphics[width=0.45\textwidth]{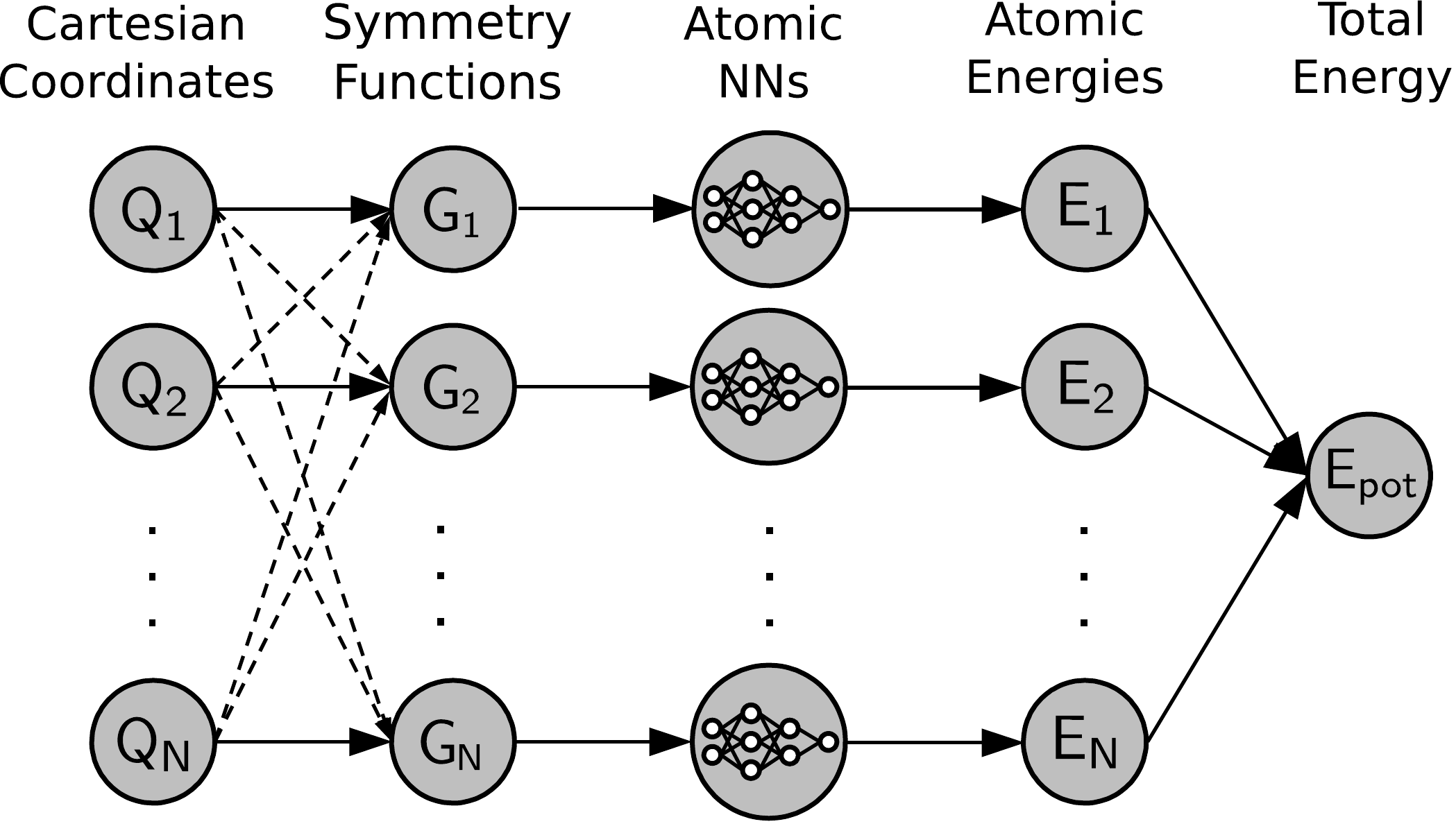}
  \caption{Schematic structure of a high-dimensional neural network potential~\cite{behler_generalized_2007}. Each Cartesian atomic coordinate vector $\mathbf{Q}_i$ is transformed to a symmetry function vector $\mathbf{G}_i$, which is used as the input for the respective atomic NN. 
The resulting energy contributions $E_i$ are summed to yield the molecule's potential energy.}
\label{fig:HDNN}
\end{figure}

Similar to their biological counterparts, NNs are assembled from several interconnected subunits, called neurons.
These neurons collect and process incoming signals (e.g. molecular geometries) and assign an output (e.g. the potential energy).
This processing is performed by computing a weighted sum and applying a nonlinear activation function, 
where the network weights control the magnitude of the incoming signals.
If the input signals are the outputs of other neurons, a network structure is obtained and the weights represent the connections between the neurons in the network.
In analogy to biological learning, the strength of these connections has to be determined in order to obtain NNs suitable for practical use and the weights are hence the important fitting parameters of a NN.\\
Unfortunately, the basic NN structure outlined above suffers from several drawbacks when applied to the interpolation of PESs. Once the weight parameters have been learned, the structure of the NN is fixed. As a consequence, the NN can only be used for molecules with the same number of atoms and elemental composition.
Moreover, the output of the NN is not invariant with respect to translations and rotations of the molecule if standard Cartesian coordinates are used as inputs.
One method to overcome these problems is the the high-dimensional NN (HDNN) approach developed by Behler and Parrinello\cite{behler_generalized_2007,behler_atom-centered_2011}.
\\
In the HDNN approach, shown schematically in Figure~\ref{fig:HDNN}, each atom of a system is characterized by its chemical environment. Depending on this environment, its energy contribution $E_i$ to the total potential energy $E_{\rm pot}$ is then calculated as output of an individual atomic NN, which is usually a conventional feed-forward NN~\cite{bishop_2006}.
By summing these energy contributions, $E_\mathrm{pot}$ is obtained. The individual atomic NNs are identical for a given element to ensure the required permutation invariance of the final PES. 

The local chemical environments of the atoms $i$ are described via sets, i.e. vectors, of many-body atom-centered symmetry functions $\mathbf{G}_i$ (ACSFs), which depend on all Cartesian atomic position vectors $\mathbf{Q}_i$ within a predefined cutoff radius around the respective central atom. 
These symmetry functions resemble radial and angular distribution functions and are invariant to translations and rotations of the molecule, thus eliminating one of the problems of standard NNs.
The introduction of a cutoff radius restricts the description of the atomic environments to the chemically relevant regions and facilitates exploiting chemical locality in the training and application of the HDNNs.
An in-depth description of HDNNs and suitable symmetry functions can be found elsewhere~\cite{Behler_Rev_2015,behler_representing_2014,behler_atom-centered_2011}.

As stated above, the weights of the NNs have to be optimized in order to obtain meaningful potential energy predictions. This is done in a process called ``training'', where a reference set of geometries and corresponding energies is iteratively reproduced to minimize the root mean squared error (RMSE) of the energies predicted by the NN.
This minimization can be achieved by a variety of algorithms, e.g. stochastic gradient descent\cite{bottou_tricks_2012} or Levenberg--Marquardt optimization\cite{levenberg_a_1944,marquart_an_1963}.
In the present work, a special adaption of the global extended Kalman filter~\cite{Kalman1960} for HDNNs, the element-decoupled Kalman filter~\cite{Gastegger2015}, has been used.
This algorithm is well suited for the flexible structure of HDNNs and results in improved training speeds and an increased quality of the resulting PESs for molecular systems.

\subsection{Systematic Fragmentation Method}

The SMF method has been used for two purposes in the present work. First its performance has been tested and compared to that of HDNNs. Second, it has been applied in combination with an energy correction scheme to provide very accurate reference energies, which enabled to test both methods for systems being inaccessible for direct coupled cluster calculations.

\begin{figure}[h]
  \centering
  \includegraphics[width=0.48\textwidth]{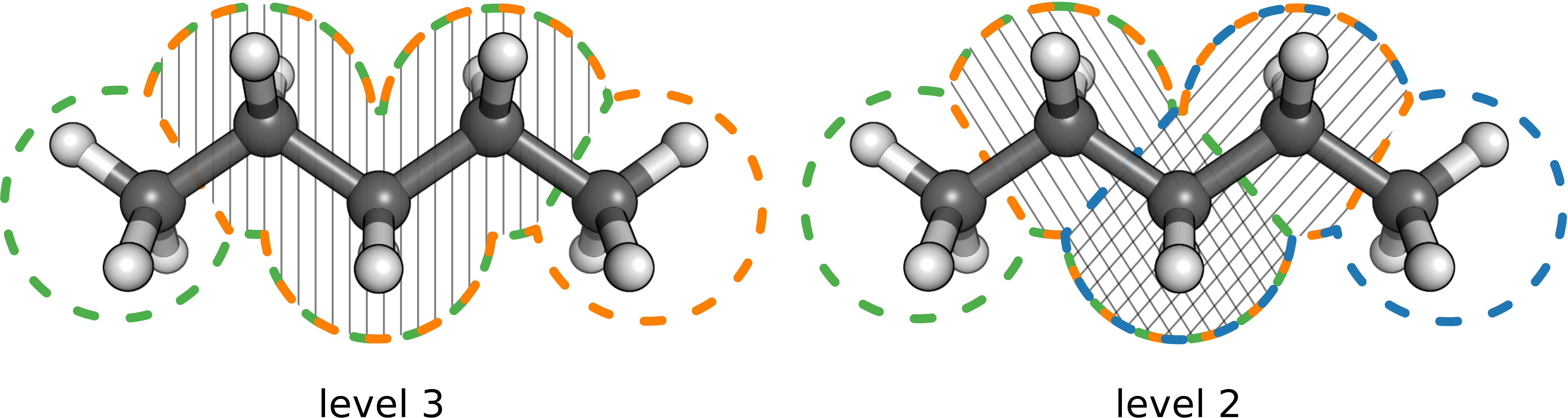}
  \caption{Exemplary fragmentation of \ce{C5H12} with fragmentation levels 3 and 2. The C-C bonds are broken homolytically, hydrogen caps are added both on the respective colored fragments and shaded overlaps. The energies of the fragments are summed and the energies of the overlaps are subtracted, yielding an approximation for the molecule's potential energy.}
  \label{fig:fragmentation}
\end{figure}

The basic principle of the SMF method \cite{Collins2006,Netzloff2007,C2CP23832B} is illustrated in Figure~\ref{fig:fragmentation} using the fragmentation of \ce{C5H12} as an example. The fragments (highlighted in green, red and blue) are constructed from the functional groups of the molecule, in the case of alkanes \ce{CH2} and \ce{CH3} groups. The single-bonds are broken homolytically and hydrogen caps are added to maintain charge neutrality. The molecule's potential energy is then approximated by adding the fragment energies and subtracting the ``double counted'', shaded overlapping regions. By using larger fragments, thus increasing the overlap size, the approximation becomes more accurate and approaches the calculation results for the entire molecule. The overlap size is denominated by the fragmentation level $X$, where $X$ indicates the number of functional groups (saturated C-atoms in our case) within the overlap.

\section{Computational Details} \label{sec:computational details}

All quantum mechanical reference calculations were carried out with ORCA~\cite{Neese2012}. Geometry optimizations were performed at the RI-MP2/cc-pVTZ level of theory~\cite{Dunning1989}. MP2 correlation and Coulomb integrals were calculated employing the resolution of identity approximation~\cite{Eichkorn1995,Vahtras1993} as well as the COSX numerical integration~\cite{Neese2009,Neese2011} for the Hartree-Fock exchange term. Single-point energies of the optimized structures and all of its fragments were obtained using implicitly correlated CCSD(T)-F12 with the resolution of identity approximation and the cc-pVTZ, cc-pVDZ-F12~\cite{Peterson2008} and cc-pVDZ-F12-CABS\cite{Yousaf2008} basis sets.

For the SMF approach, C-C bonds were broken homolytically and hydrogen caps were added according to Collins~\cite{Collins2014}, using covalent radii of 0.31~\AA\  for hydrogen and 0.76~\AA\ for carbon. In total, 9 optimized alkanes containing between 3 and 11 carbon atoms, an additional alkane with 11 carbon atoms and all 474 non-optimized fragments of these molecules have been calculated and included in the reference set irrespective of close structural similarities between many of these fragments.

The HDNN construction and training were carried out using the RuNNer code~\cite{runner}. The atomic environments were characterized by ACSFs~\cite{behler_atom-centered_2011}, whose parameters are given in the supporting information. A combination of 8 radial and 24 angular functions was employed for both carbon and hydrogen. 
A cutoff radius of 5~\AA\ was used for all ACSFs. Consequently each atomic NN contains 32 input nodes corresponding to the individual ACSFs, and one output node was used to obtain the atomic potential energy contribution. The architectures of the atomic NNs were determined by an initial training run using subnets with 1 or 2 hidden layers consisting of up to 35 nodes. Based on these preliminary training results (average error, standard deviation, minimal deviation), the five most promising architectures were chosen. The architectures are read as "first hidden layer"-"second hidden layer": 2-2, 3-4, 5-4, 10-2 and 15 for both carbon and hydrogen. Hyperbolic tangents were employed as activation function in the hidden layers, while a linear transformation was applied to the output layers. 

The training process was performed using the ``element-decoupled'' global extended Kalman filter~\cite{Gastegger2015}. The weight parameters were adjusted over 150 epochs and an adaptive filter threshold of 0.9 times the RMSE of the previous epoch was used. Values of $\lambda_0 = 0.9995$ and $\lambda_k = 0.95$ were employed for the time-dependent forgetting schedule, the network weights were initialized according to the scheme of Nguyen and Widrow~\cite{Nguyen1990}. In order to facilitate the training, the energies of the free atoms were subtracted from the reference energies of the studied molecules. Overfitting was controlled by early stopping using cross validation~\cite{behler_representing_2014} with randomly chosen training and test sets with a ratio of 9:1. Five different random seeds were used to determine training and test set compositions of each HDNN architecture. In this way, the influence of the test set composition was ensured to be negligible. The HDNN with the lowest test set RMSE was then chosen for the subsequent calculations, a model with elemental NNs of size 15 and training set and test set RMSEs of 0.00063~kcal/mol per atom and 0.00126 kcal/mol per atom respectively.

\section{Results and Discussion} \label{sec:results}

\subsection{Fragmentation}

The accuracy of the SMF approach for the model system used in this work is studied using short all-trans alkane chains with lengths ranging from 3 to 11 carbon atoms.
After geometry optimization at the MP2 level, single point energies are computed with CCSD(T). Based on these optimized geometries, systematic fragmentation is carried out with fragmentation levels from 1 to the respective maximum level given by the chain length. The energies of the full alkane molecules obtained in this way are then compared to their respective CCSD(T) values. 

\begin{figure}[ht]
  \centering
  \includegraphics[width=0.48\textwidth]{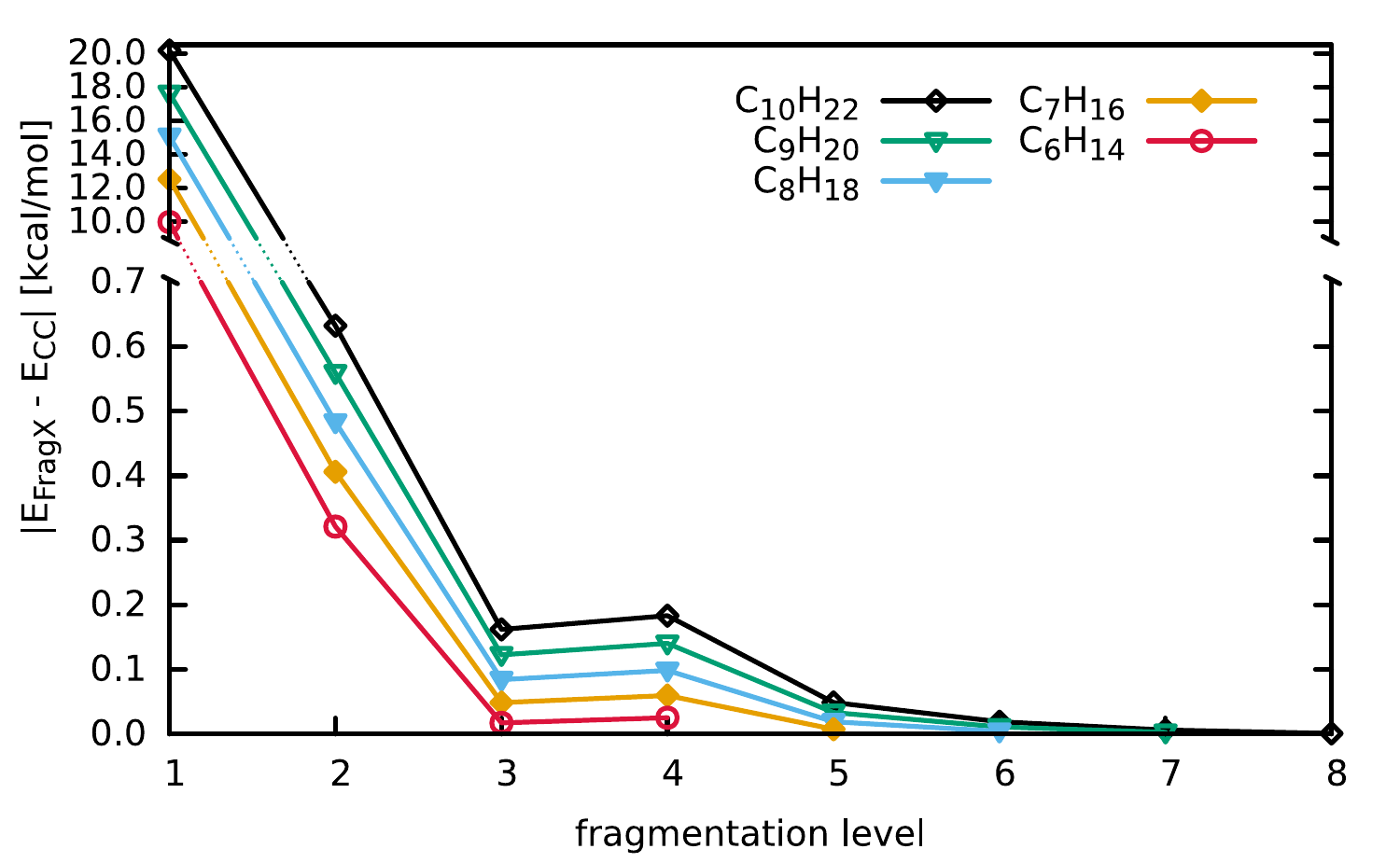}
  \caption{Total energy deviations of the energies computed at different fragmentation levels from the coupled cluster reference calculations for the alkanes \ce{C6H14} to \ce{C10H22}.}
  \label{fig:FragLevel_vs_DeltaE_merged}
\end{figure}

Using alkanes from \ce{C6H14} to \ce{C10H22} as an example, Figure~\ref{fig:FragLevel_vs_DeltaE_merged} compares the fragmentation-derived potential energies $E_\mathrm{FragX}$ with the full-sized CCSD(T) calculations $E_\mathrm{CC}$. With higher fragmentation levels the potential energy approaches the coupled cluster result of the entire molecule, as higher fragmentation levels account for a larger overlap region between the fragmentation sites. Since a higher ratio of the entire molecule is included when calculating each fragment, the general convergence trend observed in Figure~\ref{fig:FragLevel_vs_DeltaE_merged} can be expected. 
However, the increased potential energy difference of level 4 compared to 3 is interesting to note. We have not found a satisfactory explanation for this behavior.

\begin{figure}[ht]
  \centering
  \includegraphics[width=0.48\textwidth]{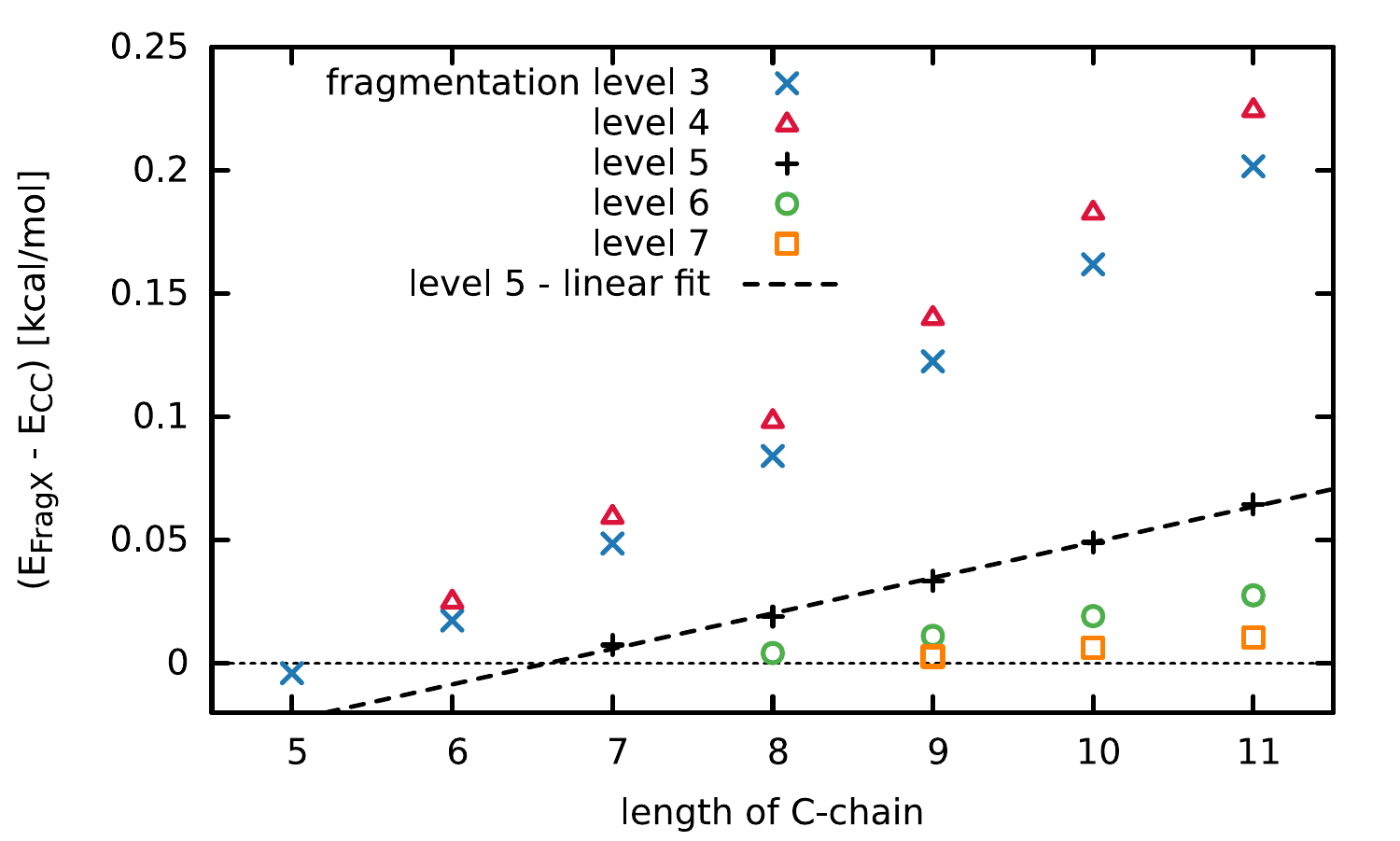}
  \caption{Energy deviations between the fragmentation method and the coupled cluster calculations for different alkanes and fragmentation levels.}
  \label{fig:ChainLength_vs_DeltaE}
\end{figure}

Figure~\ref{fig:ChainLength_vs_DeltaE} shows the deviation of fragmentation energies obtained for alkanes of length 5 to 11 using fragmentation levels starting with level 3 from the CCSD(T) results. Investigating the deviation as a function of the molecule size, the following trend can be observed: At a given fragmentation level the energy difference increases with the number of carbon atoms (a trend which can also be observed in Figure~\ref{fig:FragLevel_vs_DeltaE_merged}). The reason for this behavior is the way the energy of the whole molecule is computed in the SMF approach.
By using fragments of the same size to construct alkanes of different lengths, the corresponding error in energy is replicated with every additional C-atom, resulting in the approximately linear trend shown in Figure~\ref{fig:ChainLength_vs_DeltaE}.
Hence, the intrinsic error of the respective fragmentation level becomes visible.

This error is only small for short alkanes, but it increases with chain size.
This linear increase in the error is a consequence of the chosen model system as each \ce{CH2} group contributes approximately the same error with respect to the coupled cluster reference. This linear relation can therefore be employed to construct an energy correction for alkane chains of arbitrary length as shown for fragmentation level 5 in Figure~\ref{fig:ChainLength_vs_DeltaE}.
Taking this correction into account, the almost exact CCSD(T) values are recovered for chains with up to 11 C-atoms and it reasonable to assume that this trend holds also for longer chains, where CCSD(T) calculations are unfeasible. The corrected energies obtained by adding the correction to the fragmentation energies are denoted as $E_\mathrm{corr}$. 
While these energies can be calculated for every fragmentation level, the $E_\mathrm{corr}$ values for the different fragmentation levels show only extremely small deviations from each other (within 1.1 kcal/mol for the 10~000 carbon chain), demonstrating the stability of the correction.
In what follows, we use the $E_\mathrm{corr}$ derived from fragmentation level of 5 as it offered a sufficient amount of data points with reasonably small deviations from CCSD(T) results. 
By using this correction, we can go beyond the standard accuracy of the fragmentation method. However, this is possible only due to the linear nature of the chosen model system and such a scheme would not be applicable for arbitrary organic molecules, which is the reason why we have chosen linear alkanes for the present benchmark study.\\

\subsection{Neural Networks}

In order to assess the ability of HDNNs to model the potential energy of large linear alkanes based on the information contained in small fragments, the five NN architectures introduced in section \ref{sec:computational details}
 are used to predict potential energies of all-trans alkane chains with lengths up to 10~000 carbon atoms. 
Since the geometry optimization of alkanes of this size with \textit{ab initio} methods is impossible, model geometries are used.
These structures are obtained by replicating fragments based on the MP2 optimized bond lengths, angles and dihedral angles calculated for \ce{C10H22} until the desired length is reached. 
In the present work, alkanes containing 11 to 10~000 carbon atoms are generated in this manner.
The reference energies of these chains were computed using the level 5 fragmentation approach augmented by the previously derived correction.

The training set employed in the construction of the HDNN potentials contained the MP2 optimized alkanes (3 to 11 carbons) and a \ce{C11H24} structure generated as outlined in the previous paragraph, as well as all their respective fragments.
Note that a training set with purely MP2 based geometries leads to strong fluctuations in the predicted potential energies for the long artificial chains.
The reason for this effect is the highly regular nature of the linear alkane model system, which prevents a comprehensive sampling of the possible configuration space. As a consequence, the HDNNs are sensitive with respect to tiny differences between MP2-optimized and artificially generated geometries, which is not expected to happen in typical molecular systems if the PESs are based on more representative data sets of the relevant configuration space.

\begin{figure}[ht]
  \centering
  \includegraphics[width=0.48\textwidth]{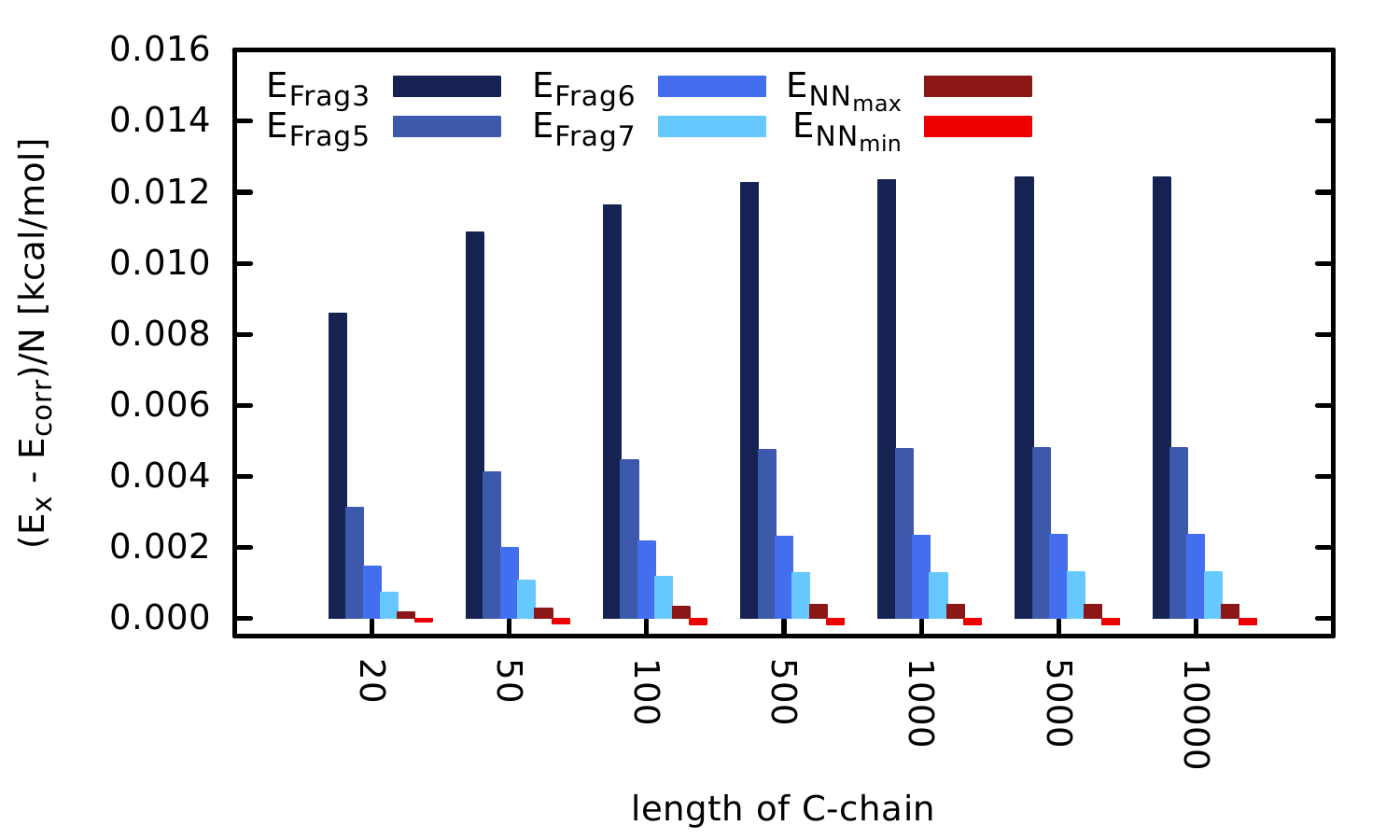}
  \caption{Potential energies derived from fragmentation and NN approximations ($E_\mathrm{FragX}$ and $E_\mathrm{NN}$)  in comparison to the error-corrected level 5 fragmentation results $E_\mathrm{corr}$. The energy is normalized to the number of atoms $N$ of the alkanes.}
  \label{fig:E_x_minus_E_frag5corrected_extract}
\end{figure}

The deviations from $E_\mathrm{corr}$ of the potential energies computed with the different methods are illustrated in Figure~\ref{fig:E_x_minus_E_frag5corrected_extract}. For the HDNNs, the predictions with the highest and lowest deviation ($E_\mathrm{NN_{max}}$ and $E_\mathrm{NN_{min}}$) are shown. They are compared with the $E_\mathrm{FragX}$ approximations, where $X$ denominates the corresponding fragmentation level. In order to achieve a reasonable scale, the energy is normalized to the number of atoms $N$ for demonstrative purposes.

Once again, the trend of the $E_\mathrm{FragX}$ energies to yield more accurate approximations with higher fragmentation levels can be observed.
However, all HDNN approximations yield even smaller deviations from $E_\mathrm{corr}$, with the maximum deviations still lying below the ones obtained for fragmentation level 7 and the best HDNNs ($E_\mathrm{NN_{min}}$) performing significantly better. This result is remarkable insofar, as the choice of a 5~\AA\ cutoff radius used in the ACSFs limits the effective chemical environment seen by a HDNN to a maximum of 7 carbon atoms. Compared to the SMF method, this number of carbons corresponds to a fragmentation level of 6, which shows significantly larger deviations than the HDNNs.
Apparently, the HDNNs are able to exploit chemical locality to a greater extent compared to the standard SMF method and hence utilize the information present in the molecular fragments in a more efficient manner.

The errors in the NN predictions in general exhibit the same linearity as the fragmentation derived values, which is to be expected as also for the HDNN potential each additional CH$_2$ group contributes a certain energy error, but for the given reference the NN energies are notably more accurate. Thus, HDNNs represent a promising alternative to the fragmentation method, and we believe that this finding also holds for general organic molecules. A comparison between the SMF approach and HDNNs is more difficult in this case, as no simple corrections can be exploited and accurate reference data is hence more difficult or even impossible to obtain.

\section{Conclusion}

A comparison of the performance of high-dimensional neural network (HDNNs) potentials and of the SMF approach for the energies of linear all-trans alkanes has been presented. Due to the linearity of the energy error of the fragmentation approach with system size an energy correction scheme could be implemented that enabled to assess the accuracy of both methods for systems containing up to 10~000 C-atoms. While both approaches provide very accurate energies close to the underlying coupled cluster data, the energy errors employing the HDNN approach have been found to be systematically smaller for all chain lengths.
Unlike the fragmentation method, the purely mathematical structure of HDNNs is not restricted by underlying physical considerations.
Another advantage of HDNN potentials is their transferability. Once trained, they can be used to obtain the energy of sufficiently similar molecules, without the need of additional \textit{ab initio} calculations.
This principal flexibility, accuracy and efficiency illustrates the benefits of HDNNs for other chemical systems and applications. 
However, it should once again be stressed, that the model system studied in this work is extremely well behaved and exhibits no significant long range electrostatic or dispersion interactions.
Whether the results of our particular model system can be reproduced for more complex systems like proteins will be subject of further studies.

\section*{Supplementary Material}

See supplementary material for a listing of the symmetry functions and their respective parameters used to describe the local chemical environments in the present work.

\section*{Acknowledgement}
Allocation of computer time at the Vienna Scientific Cluster (VSC) is gratefully acknowledged.
JB is grateful for financial support by the DFG.

\newpage
~
\newpage

\section*{Supplementary Material: Comparing the Accuracy of High-Dimensional Neural Network Potentials and the Systematic Molecular Fragmentation Method: A Benchmark Study for all-trans Alkanes}

\section{Symmetry Function Parameters}

The local chemical environment of the different atoms in the all-trans alkanes is characterized via 8 radial symmetry functions of the type
\begin{equation}
G^{\mathrm{rad}}_{i} = \sum^{N_{\mathrm{atoms}}}_{j \neq i} e^{-\eta (R_{ij}-R_\textrm{s})^{2}} f_{\mathrm{c}}(R_{ij}), \label{eq:rad}
\end{equation}
and 24 angular symmetry functions 
\begin{align}
G^{\mathrm{ang}}_{i} = 2^{1-\zeta} \sum^{N_{\mathrm{atoms}}}_{j,k \neq i} & \left( 1 + \lambda \theta_{ijk}\right) e^{-\eta (R^{2}_{ij} + R^{2}_{ik} + R^{2}_{jk} ) } \nonumber \\
& \times f_{\mathrm{c}}(R_{ij}) f_{\mathrm{c}}(R_{ik}) f_{\mathrm{c}}(R_{jk}). \label{eq:ang}
\end{align}
$R_{ij}$ is the distance between atoms $i$ and $j$ (analogous also for atoms $k$), $R_s$ is the offset of the Gaussian function. $\eta$, $\zeta$ and $\lambda$ are parameters which determine the overall shape of the symmetry functions. $f_c$ is a cutoff function introduced to limit the description of the local environment to the chemically relevant regions and is defined as

\begin{equation*}
f_{\textrm{c}}(R_{ij}) = 
\begin{cases}
\frac{1}{2} \left[ \cos \left( \frac{\pi R_{ij}}{R_{\mathrm{c}}} \right) + 1 \right],&  R_{ij} \leq R_{\textrm{c}}\\
                                                                                   0,&  R_{ij} > R_{\textrm{c}},
\end{cases}
\end{equation*}
with $R_c$ as the cutoff radius. For a more detailed discussion of the different symmetry functions, see Reference~\citenum{behler_atom-centered_2011}.

The parameters of the radial and angular symmetry functions used to describe the environment of hydrogen atoms in the linear all-trans alkanes are given in Table~\ref{tab:Hrad} and Table~\ref{tab:Hang}.
Those defining the symmetry functions of carbon are given in Tables~\ref{tab:Crad} and \ref{tab:Cang}.


%\newpage
%~
%\newpage

% H radial
\begin{table}[!htbp]
\begin{center}
\caption{Parameters of the radial symmetry functions (Eqn.~\ref{eq:rad}) describing the chemical environment of H atoms.}
\label{tab:Hrad}
\begin{tabular}{ccrrr}
\toprule
No. & Neighbor & $\eta$ [Bohr$^{-2}$] & $R_s$ [Bohr] & $R_c$ [Bohr]\\
\hline
1 & H  & 0.00558645 &     0.0 &  9.44865 \\
2 & H  & 0.01117290 &     0.0 &  9.44865 \\
3 & H  & 0.02235000 &     0.0 &  9.44865 \\
4 & H  & 0.04469000 &     0.0 &  9.44865 \\
5 & C  & 0.01587000 &     0.0 &  9.44865 \\
6 & C  & 0.03175000 &     0.0 &  9.44865 \\
7 & C  & 0.06350000 &     0.0 &  9.44865 \\
8 & C  & 0.12700000 &     0.0 &  9.44865 \\
\hline
\end{tabular}
\end{center}
\end{table}
% H angular
\begin{table}[!htbp]
\begin{center}
\caption{Parameters of the angular symmetry functions (Eqn.~\ref{eq:ang}) describing the chemical environment of H atoms.}
\label{tab:Hang}
\begin{tabular}{ccrrrr}
\toprule
No. & Neighbors & $\eta$ [Bohr$^{-2}$] & $\lambda$ & $\zeta$ & $R_c$ [Bohr]\\
\hline
~9 & H H & 0.00558645 &  1.0  & 4.0 &  9.44865 \\
10 & H H & 0.01117290 &  1.0  & 1.0 &  9.44865 \\
11 & H H & 0.02235000 &  1.0  & 1.0 &  9.44865 \\
12 & H H & 0.04469000 &  1.0  & 1.0 &  9.44865 \\
13 & H H & 0.00558645 & -1.0  & 4.0 &  9.44865 \\
14 & H H & 0.01117290 & -1.0  & 1.0 &  9.44865 \\
15 & H H & 0.02235000 & -1.0  & 1.0 &  9.44865 \\
16 & H H & 0.04469000 & -1.0  & 1.0 &  9.44865 \\
\hline
17 & C H & 0.01587000 &  1.0  & 4.0 &  9.44865 \\
18 & C H & 0.03175000 &  1.0  & 1.0 &  9.44865 \\
19 & C H & 0.06350000 &  1.0  & 1.0 &  9.44865 \\
20 & C H & 0.12700000 &  1.0  & 1.0 &  9.44865 \\
21 & C H & 0.01587000 & -1.0  & 4.0 &  9.44865 \\
22 & C H & 0.03175000 & -1.0  & 1.0 &  9.44865 \\
23 & C H & 0.06350000 & -1.0  & 1.0 &  9.44865 \\
24 & C H & 0.12700000 & -1.0  & 1.0 &  9.44865 \\
\hline
25 & C C & 0.00074765 &  1.0  & 4.0 &  9.44865 \\
26 & C C & 0.01495304 &  1.0  & 1.0 &  9.44865 \\
27 & C C & 0.02991000 &  1.0  & 1.0 &  9.44865 \\
28 & C C & 0.05981000 &  1.0  & 1.0 &  9.44865 \\
29 & C C & 0.00074765 & -1.0  & 4.0 &  9.44865 \\
30 & C C & 0.01495304 & -1.0  & 1.0 &  9.44865 \\
31 & C C & 0.02991000 & -1.0  & 1.0 &  9.44865 \\
32 & C C & 0.05981000 & -1.0  & 1.0 &  9.44865 \\
\hline   
\end{tabular}
\end{center}
\end{table}

\newpage

% C radial
\begin{table}[!htbp]
\begin{center}
\caption{Parameters of the radial symmetry functions (Eqn.~\ref{eq:rad}) describing the chemical environment of C atoms.}
\label{tab:Crad}
\begin{tabular}{ccrrr}
\toprule
No. & Neighbor & $\eta$ [Bohr$^{-2}$] & $R_s$ [Bohr] & $R_c$ [Bohr]\\
\hline
1 & H  & 0.01587000 &     0.0 &  9.44865 \\
2 & H  & 0.03175000 &     0.0 &  9.44865 \\
3 & H  & 0.06350000 &     0.0 &  9.44865 \\
4 & H  & 0.12700000 &     0.0 &  9.44865 \\
5 & C  & 0.00074765 &     0.0 &  9.44865 \\
6 & C  & 0.01495304 &     0.0 &  9.44865 \\
7 & C  & 0.02991000 &     0.0 &  9.44865 \\
8 & C  & 0.05981000 &     0.0 &  9.44865 \\
\hline   
\end{tabular}
\end{center}
\end{table}
% C angular
\begin{table}[!htbp]
\begin{center}
\caption{Parameters of the angular symmetry functions (Eqn.~\ref{eq:ang}) describing the chemical environment of C atoms.}
\label{tab:Cang}
\begin{tabular}{ccrrrr}
\toprule
No. & Neighbors & $\eta$ [Bohr$^{-2}$] & $\lambda$ & $\zeta$ & $R_c$ [Bohr]\\
\hline
~9 & H H & 0.00558645 &  1.0  & 4.0 &  9.44865 \\
10 & H H & 0.01117290 &  1.0  & 1.0 &  9.44865 \\
11 & H H & 0.02235000 &  1.0  & 1.0 &  9.44865 \\
12 & H H & 0.04469000 &  1.0  & 1.0 &  9.44865 \\
13 & H H & 0.00558645 & -1.0  & 4.0 &  9.44865 \\
14 & H H & 0.01117290 & -1.0  & 1.0 &  9.44865 \\
15 & H H & 0.02235000 & -1.0  & 1.0 &  9.44865 \\
16 & H H & 0.04469000 & -1.0  & 1.0 &  9.44865 \\
\hline
17 & C H & 0.01587000 &  1.0  & 4.0 &  9.44865 \\
18 & C H & 0.03175000 &  1.0  & 1.0 &  9.44865 \\
19 & C H & 0.06350000 &  1.0  & 1.0 &  9.44865 \\
20 & C H & 0.12700000 &  1.0  & 1.0 &  9.44865 \\
21 & C H & 0.01587000 & -1.0  & 4.0 &  9.44865 \\
22 & C H & 0.03175000 & -1.0  & 1.0 &  9.44865 \\
23 & C H & 0.06350000 & -1.0  & 1.0 &  9.44865 \\
24 & C H & 0.12700000 & -1.0  & 1.0 &  9.44865 \\
\hline
25 & C C & 0.00074765 &  1.0  & 4.0 &  9.44865 \\
26 & C C & 0.01495304 &  1.0  & 1.0 &  9.44865 \\
27 & C C & 0.02991000 &  1.0  & 1.0 &  9.44865 \\
28 & C C & 0.05981000 &  1.0  & 1.0 &  9.44865 \\
29 & C C & 0.00074765 & -1.0  & 4.0 &  9.44865 \\
30 & C C & 0.01495304 & -1.0  & 1.0 &  9.44865 \\
31 & C C & 0.02991000 & -1.0  & 1.0 &  9.44865 \\
32 & C C & 0.05981000 & -1.0  & 1.0 &  9.44865 \\
\hline   
\end{tabular}
\end{center}
\end{table}

\end{document}